# Opening of large band gap in metallic carbon nanotubes by mannose functionalized dendrimers: Experiments and theory

K. S. Vasu[a*], Debabrata Pramanik[a], Sudipta Kundu[a], Sridevi S[b], N. Jayaraman[c], Manish Jain[a], Prabal K. Maiti[a] and A. K. Sood[a]

[a]Department of Physics, Indian Institute of Science, Bangalore-560012, India.

[b]Department of Instrumentation and Applied Physics, Indian Institute of Science, Bangalore-560012, India.

[c]Department of Organic Chemistry, Indian Institute of Science, Bangalore-560012, India.

**Keywords:** Carbon nanotube, Dendrimer, Metal-semiconductor transition, Band gap opening, Ionic liquid top gating, Density functional theory, GW calculations.

**Abstract**

Despite many theoretical schemes, the direct experimental observation of supramolecular control on band gap opening in single-walled carbon nanotubes (SWNT) is still lacking. We report experimental and theoretical demonstration of metal to semiconductor transition with precisely measured large band gap in SWNTs due to the wrapping of mannose functionalized poly (propyl ether imine) dendrimer (DM) molecules. Semiconductor behaviour of SWNT-DM complex is comprehensively established with a band gap value of ~ 0.45 eV measured using scanning tunnelling spectroscopy (STS), ionic liquid top gated field effect transistor (FET) characteristics and Raman spectroscopy. Further, a validated molecular picture of SWNT-DM complex obtained from fully atomistic molecular dynamic (MD) simulations was used to carry out *ab-initio* density functional theory (DFT) and GW calculations of the electronic structure, evaluating experimentally estimated band gap value. We attribute this large band gap opening in SWNTs to the complexation induced asymmetric strain developed in the carbon-carbon bond length.

**Introduction:**

Exceptional electronic and mechanical properties of carbon nanotubes made them promising candidates for the fabrication of nanoelectronic devices[1-4]. Geometry (diameter and chirality) dependant translational, rotational and helical symmetries in SWNTs have a strong influence on their electronic properties which define the semiconducting and metallic behaviour[1]. However, tuning the electronic properties of SWNTs by precisely controllable chemical fucntionalization[5-8] broadens their potential applications in nanofabrication and biology[9-13]. Heretofore, band gap opening and tuning the electronic behaviour of SWNTs have been intensively discussed in theoretical literature[14-19], but with, only a little known input of molecular functionalization from experimental perspective[5-7,20]. So far in experiments, DNA, viologens, gold and $Cu_2O$ nanoparticles are the only molecules which exhibited the ability to open a band gap in SWNTs. Moreover, these studies report purely a qualitative demonstration of very small band gap opening in SWNTs with barely a discussion about its quantification. Thus, the discovery of novel molecular structures for tuning the SWNT's electronic properties and the precise quantification of induced band gap value with theoretical realization are of demanding interest, due to the significance in achieving an efficient utilization of SWNTs in nanofabrication[9] and biological applications[11]. Here, we report a large band gap (~ 0.45 eV) opening in SWNTs due to wrapping with mannose functionalized fourth generation (G4) oxygen core poly (propyl ether imine) dendrimer (DM) molecule (Fig. S1), a synthetic macromolecule with branched structure[3]. We have precisely estimated the band gap value (0.45 eV) using ionic liquid top gated FET characteristics and STS measurements. Using all atom MD simulations along with DFT and GW calculations of the electronic DOS, we attribute this large band gap opening in metallic SWNTs to the asymmetric strain developed in the carbon-carbon bond length. Unlike the SWNT-ssDNA complex, surrounding water molecules do not have any role in metal to semiconductor transition in SWNT-DM complex.

**Experimental**

**Device fabrication**

Aqueous suspensions of metallic SWNTs (concentration ~ 0.1 mg/mL and average diameter ~ 1.4 nm) obtained from M/s NanoIntegris and mannose functionalized fourth generation (G4) oxygen core dendrimer synthesized by us were used in the present study. Fig. 1(a) shows schematic of the device and the scanning electron microscope (SEM) image of metallic SWNTs decorated with the DM molecules (SWNT-DM complex). Fig. S1 shows the chemical structure of the DM molecule. SWNT FET devices were fabricated using ac dielectrophoresis of aqueous suspensions of metallic SWNTs on pre-patterned Cr/Au electrodes of width 5 μm and channel length 1 μm[3,21]. 5 μL of aqueous SWNT suspension (10 ng/mL) was dropped between the electrodes and an ac signal with amplitude 10 V peak to peak at 5 MHz was applied for 2 mins. Subsequently, the droplet was blown away and FET device was washed in methanol and water followed by heating at 70°C for 15 mins. Functionalization of DM molecules was carried out by placing 5 μL of DM solution (0.5 mM) onto the SWNT FET device for 15 mins[3]. After this process, the device was thoroughly washed in water for 5 mins.

**Ionic liquid gating measurements**

Electrical characterization of the FET devices was carried out using Keithley 2400 source meters. Ionic electrolyte 1-ethyl-3-methylimidazolium bis (trifluoromethylsulfonyl) imide (EMIM TFSI) was used for electrochemical top gating of the FET device made of SWNT-DM complex. EMIM TFSI ionic liquid was stored in vacuum (~ $10^{-6}$ Torr) for 12 hours prior to its use for gating experiments. Ionic liquid gating experiments were carried out in vacuum (~ $3 \times 10^{-3}$ Torr) by carefully placing 1 µL of EMIM TFSI solution onto the FET device.

**Raman and STS characterization**

Micro Raman spectra of metallic SWNTs and SWNT-DM complex were recorded using LabRAM HR Raman spectrometer (HORIBA scientific) with 633 nm excitation (laser power < 3 mW). Room temperature STM and STS measurements experiments were carried out using Bruker Innova atomic force microscope (AFM). For STM imaging of pristine metallic SWNTs and SWNT-DM complex, the instrument was set to constant current mode with an applied bias of 0.1 V to the sample and tunnelling current of 1 nA. The samples for STM and STS measurements were prepared by vacuum drying of 50 µL of aqueous metallic SWNTs suspension (1 ng/ml) on HOPG followed by methanol and water washing for 30 mins. 100 µL of DM solution (0.5 µM) was then added to metallic SWNTs on the HOPG and waited for 30 mins to prepare the SWNT-DM complex. This sample was then washed with water for 15 mins and dried in vacuum for 2 hours prior to STM imaging and STS measurements.

Detailed computational methods used for DFT and GW calculations are provided in the supplementary information.

**Results and discussion**

**Raman spectroscopy**

Raman spectroscopy, an effective and established technique for distinguishing the metallic and semi-conducting SWNTs, was used to confirm the metal to semiconductor transition in SWNTs. Fig. 1(b) shows Raman spectrum corresponding to the tangential G band and the inset shows radial breathing mode (RBM) of pristine metallic SWNTs and SWNT-DM complex. The band center ($\omega_{RBM}$) of dispersive RBM of metallic SWNTs (centered ~ 172 cm$^{-1}$) and SWNT-DM complex (centered ~ 170 cm$^{-1}$) did not show any significant change. We have estimated the diameter of SWNTs as ~ 1.4 nm from the relation between $\omega_{RBM}$ and diameter of the tubes[1] (see Fig. S2 for additional Raman characteristics). As seen in Fig. 1b, tangential G band of metallic SWNTs has shown substantial changes in both the shape and Raman shift before and after functionalization with the DM molecules. A strong BWF component centered at ~ 1542 cm$^{-1}$ (G$^-$ band) and a Lorentzian component centered at ~ 1584 cm$^{-1}$ (G$^+$ band) are observed in G band of pristine metallic SWNTs. After functionalization of metallic SWNTs with the DM molecules, the SWNT-DM complex has shown G-band, two Lorentzians centered at ~ 1546 cm$^{-1}$ (G$^-$ band) and ~ 1592 cm$^{-1}$ (G$^+$ band), similar to that of typical semiconducting SWNTs. The BWF component in Raman spectrum of metallic SWNTs appears due to the strong coupling between tangential vibrational modes and conduction electrons, not present in the case of semiconducting tubes[22]. Therefore, the absence of BWF component and the blue shift in high frequency G$^+$ band of metallic SWNTs after functionalization confirm the semiconducting nature of the SWNT-DM complex.

**Back gate transfer characteristics**

Inset of Fig. 2(a) shows source-drain current-voltage ($I_{DS}$-$V_{DS}$) and the main panel shows back gate transfer ($I_{DS}$-$V_{BG}$) characteristics of the device made of the metallic SWNTs before and after functionalization with the DM molecules. Typical ohmic $I_{DS}$-$V_{DS}$ characteristics (inset of Fig. 2(a)) of the FET device made of metallic SWNTs acquired non-ohmic component with an increased resistance value after the functionalization with the DM molecules. As expected, the

FET device made of metallic SWNTs did not show any gating response during the entire $V_{BG}$ sweep from +50 V to -20 V, $I_{DS}$-$V_{BG}$ characteristics recorded at constant $V_{DS}$ = 0.1 V. However, the same device has shown p-type semiconducting $I_{DS}$-$V_{BG}$ characteristics with the onset in $I_{DS}$ at ~ -5.5 V, after functionalization. The p-type behaviour of metallic SWNT FET device after functionalization suggests the electron transfer from the metallic nanotubes to the DM molecules. The absence of ambipolar transfer characteristics even at $V_{BG}$ = +50 V can be attributed to asymmetric Schottky barrier formation for the charge carriers at electrode interfaces. Since the experiment is performed at room temperature under ambient conditions, reverse run from -20 V to +50 V has shown a large hysteresis due to the charge trapping by water molecules[23] around the SWNTs.

**Top gate transfer characteristics**

We have also carried out electrolytic top gating using EMIM TFSI ionic liquid to achieve ambipolar transfer characteristics for quantitative estimation of the band gap opening[24]. Since the ionic liquids form high dielectric and very thin (~ 1 nm) electric double layers, we can achieve high carrier density into the semiconducting channel even at very small gate voltages. Fig. 2(b) shows $I_{DS}$ as a function of top gate voltage ($V_{TG}$) of the FET device made of SWNT-DM complex. $V_{TG}$ was ramped from -0.5 V to 1 V in a step size of 0.025 V with waiting time of 100 s at each gate voltage. The inset of Fig. 2(b) shows the derivative of $I_{DS}$ as a function of $V_{TG}$. From the log scale plot of $I_{DS}$ as a function of $V_{TG}$, the estimated sub-threshold swing value of our FET device is ~ 250 mV/decade. The difference in onset of increase in $I_{DS}$ for the hole doping (at -0.05 V) and electron doping (at 0.4 V) supports our previous suggestions of asymmetric Schottky barrier formation at the electrode interfaces. The very small OFF-state current of ~ 60 nA of the FET device made of SWNT-DM complex suggests the negligible density of electronic states present in the band gap which is confirmed later by the STM measurements and electronic DOS calculations.

The applied $V_{TG}$ creates an electrostatic potential difference '$\phi$' between the SWNTs and gate electrode and a shift in the Fermi level ($E_F$) due to doping of carriers. Thus the change in $V_{TG}$ can be written as $\Delta V_{TG} = (\Delta E_F/e) + \Delta\phi$ with $\Delta E_F/e$ being determined by quantum capacitance. $\Delta\phi = ne/C_{TG}$, where $C_{TG}$ is the geometrical capacitance and 'n' is the carrier concentration given by $\int D(\varepsilon)f(\varepsilon)d\varepsilon$; $D(\varepsilon)$ is the density of states and $f(\varepsilon)$ is the Fermi-Dirac distribution. The typical value of $C_{TG}$ for EMIM TFSI reported from ac impedance technique is ~ 10 µF/cm$^2$ (at 0.1 Hz) and ~ 7 µF/cm$^2$ (at 1 Hz)[25] which is ~ 800 times higher than the gate capacitance of 300 nm SiO$_2$. Considering the high $C_{TG}$ value, small OFF-state current and low carrier concentration due to the absence of doping, $\Delta\phi$ is negligible and $e\Delta V_{TG} \approx \Delta E_F$ in OFF-state of the FET device made of SWNT-DM complex. Thus, we are justified in assuming the applied $V_{TG}$ is entirely utilized for shifting the Fermi level between valence and conduction bands. Therefore, the estimated transport gap (where $dI_{DS}/dV_{TG} = 0$) is ~ 0.45 eV.

**STM and STS measurements**

Further quantification of band gap opening was accomplished by low current STM and STS measurements of metallic SWNTs and SWNT-DM complex. Since the experiments were performed at atmospheric pressure, bias voltage ($V_{bias}$) applied to the sample is limited from -1 V to +1 V to operate the STM instrument in tunnelling regime. First, we have performed STM imaging to find individual SWNTs as well as SWNT-DM complex on HOPG. Subsequently, the tunnelling current ($I_t$) was recorded from pristine metallic SWNTs and SWNT-DM complex as a function of $V_{bias}$ in the STS measurements. To validate the metal to semiconductor transition and quantify the band gap value, we then estimated the local dynamic tunnelling conductance ($dI_t/dV_{bias}$) which represents the local DOS, as a function of $V_{bias}$ from STS measurements. Fig. S3(a) and the inset show the STM image of metallic SWNTs deposited on a HOPG substrate and the height profile, confirming the diameter of SWNTs to be ~ 1.4 nm, in agreement with Raman spectroscopy. $dI_t/dV_{bias}$ vs $V_{bias}$ estimated for metallic SWNTs (Fig. S3(b)) from the STS measurements has shown significant and finite value at all the bias voltage from -1 V to +1 V, confirming the absence of band gap. The measured value of $\Delta E^M_{11}$ ~ 1.4 eV between the first van Hove singularities is in agreement with the calculated value for a metallic tube of diameter ~1.4 nm.

Fig. 2(c) shows the STM topographical image of SWNT-DM complex deposited on a HOPG substrate. The red colour elliptical rings denote the DM molecules wrapped around SWNTs. Fig. 2(d) shows $I_t$ vs $V_{bias}$ (Inset) and $dI_t/dV_{bias}$ vs $V_{bias}$ collected from the marked areas for the SWNT-DM complex. $dI_t/dV_{bias}$ ~ 0 between $V_{bias}$ of -0.22 V and +0.23 V is due to the absence of electronic DOS and hence represents the band gap region for the SWNT-DM complex. The tunnelling process during the positive and negative $V_{bias}$ values is schematically shown in Fig. S4. When $V_{bias}$ = +0.23 V, electrons tunnel from the metallic tip into the first van Hove singularity in the conduction band of the SWNT-DM complex. On the other hand, when $V_{bias}$ = -0.22 V, electrons transfer from the complex into the unoccupied states of the tip. Therefore, the band gap value of the SWNT-DM complex ~ 0.45 eV from the STS measurements is in good agreement with the transport gap (~ 0.45 eV) measured in top gated FET characteristics (Fig. 2b).

We further show that, unlike the SWNT-ssDNA complex[6], the presence or absence of water molecules does not affect the wrapping strength, charge transfer process and stability of SWNT-DM complex. Fig. S5 shows that no change is observed in $I_{DS}$ -$V_{BG}$ characteristics of the device made of SWNT-DM complex before (black colour solid line) and after vacuum drying for 24 hours (red colour dashed line). Numerous alkyl chains present in the DM molecule provide the stable van der Waals interactions with the SWNT. $I_{DS}$ value (at $V_{BG} = 0$) of the same device is increased when temperature is raised from 300 K to 325 K, which further confirming the semiconducting behaviour of the SWNT-DM complex.

**DFT and GW calculations**

To elucidate the physical origin and mechanism of metal to semiconductor transition, we first obtained molecular level equilibrium microscopic picture of the SWNT-DM complex from all atom MD simulations and carried out DFT calculations for electronic DOS. We have considered all three types of nanotubes with the same diameter (∼ 1.4 nm): zigzag (18,0), chiral (12,9) and armchair (10,10). Due to strong van der Waals attraction, the DM and Poly (propyl ether imine) (PETIM) molecules get adsorbed onto the SWNT surface and form a stable composite structure (see Fig. S6(a), (b), (c) and (d)). Once the atomic level microscopic picture of SWNT binding with DM and PETIM molecules is obtained, we take the modified coordinates of the carbon atoms in the nanotube and calculate its electronic structure. As we will quantify later, the wrapping of the DM induces anisotropic strain (different along and perpendicular to the nanotube axis) in the nanotube

Fig. 3(a) shows the electronic DOS for (18,0) metallic SWNT for two different cases: when SWNT is wrapped by the DM molecule (shown in red) and when SWNT is wrapped by PETIM dendrimer without Mannose end groups (shown in cyan). DOS calculations were carried out by considering the strained SWNT (obtained from MD simulations) and each of this DOS is obtained by averaging over many independent initial configurations. It is noteworthy that a band gap of ~ 0.05 eV is observed near Fermi level in the case of SWNT wrapped with DM molecules, whereas only a dip in the DOS for the SWNT wrapped with the PETIM dendrimer. Though the band gap opening in the SWNT-DM complex is realized from DFT, the calculated gap is much smaller than the experimental value. To obtain a better estimate of the band gap, we performed GW calculations on one of the nanotube snapshot from SWNT-DM complex. Fig. 3(c) and 3(d) show the electronic band structure of the pristine (18,0) SWNT and for the SWNT-DM complex. Band structure of the (18,0) SWNT and DM wrapped (18,0) SWNT is calculated within GW (red solid line) as well as DFT (blue dotted line), setting the valence band maximum (VBM) at zero for both of them (also see Fig. S7). The pristine (18,0) SWNT is metallic in both DFT and GW. The DM wrapped SWNT shows a clear increase in the band gap in comparison to DFT. The GW calculation on the DM wrapped tube yields a band gap of 0.49 eV. This is in excellent agreement with the experimental values (0.45 eV).

Is this band gap opening unique to (18,0) chirality, or can one expect such band gap opening in metallic SWNT with the same diameter but different chirality? To answer this question, we have also looked at the morphology of the DM wrapped metallic SWNT having (10,10) and (12,9) chirality and calculate the electronic DOS as shown in Fig. 3(b). It's clear from the DOS that there is no band gap for (10,10) and (12,9) SWNT as compared to (18,0) by the DM wrapping. The plot of DOS in Fig. 3(b), uses a Gaussian broadening of 25 meV. As a result of this broadening and averaging over configurations, even though some of the tubes are semiconducting, the average DOS appears metallic. For the (10,10) tube with a small gap, we further performed GW calculation. In order to obtain the converged GW calculation for this SWNT, we had to increase its k-point sampling to 1x1x24. Within GW, the gap opens from 16.4 meV to 64.2 meV. This value is very small as compared to the experimental gap, indicating that a large bandgap opening is unique to (18,0) tubes.

Our DFT calculations show that there is no charge transfer between the SWNT and DM molecules as evident from the Bader charge analysis (see Fig. S8). Furthermore, the fat band analysis shows that the states near the Fermi level retain their π character in the DM wrapped SWNT (see Fig. S9 and S10). To understand the physical origin of this metal to semiconducting transition, we calculate two types of bond lengths 'a' and 'b' (as shown in Fig. 4) for the strained SWNT. We find that for the SWNT-DM (18,0), the bond asymmetry (a-b) is almost twice as compared to the bond asymmetry for the SWNT-PETIM (18,0) case. For the SWNT-DM (10,10) and SWNT-DM (12,9), the bond asymmetry is even less. We therefore attribute the higher bond asymmetry in (18,0) SWNT due to the wrapping of DM as the origin for metal to semiconductor transition.

**Conclusions**

In summary, we have demonstrated metal to semiconductor transition in metallic SWNTs and precisely measured the band gap value (~ 0.45 eV) using both experiments and theory. The FET experiments (see Fig. S11) on metallic SWNTs wrapped with PETIM dendrimer without mannose functionality did not show metal to semiconductor transition, which was also confirmed from the DOS calculations. We can thus suggest that $((CH_2O)_6\text{-O-N-})$ in the DM molecule is necessary to induce bond length asymmetry and hence opening up of the band gap. Our work is an experimental and theoretical effort for explicit understanding of the supramolecular control of band gap opening in SWNTs. SWNT-DM complexes exhibit wide range of biosensing applications and have the potential for fabricating spatially patterned metallic SWNTs with nm periodicity to produce efficient photodetectors.

**Acknowledgement**

We thank Department of Science and Technology for financial assistance under the Nanomission grant. AKS thanks Year of Science Fellowship for the financial assistance.

**Figures:**

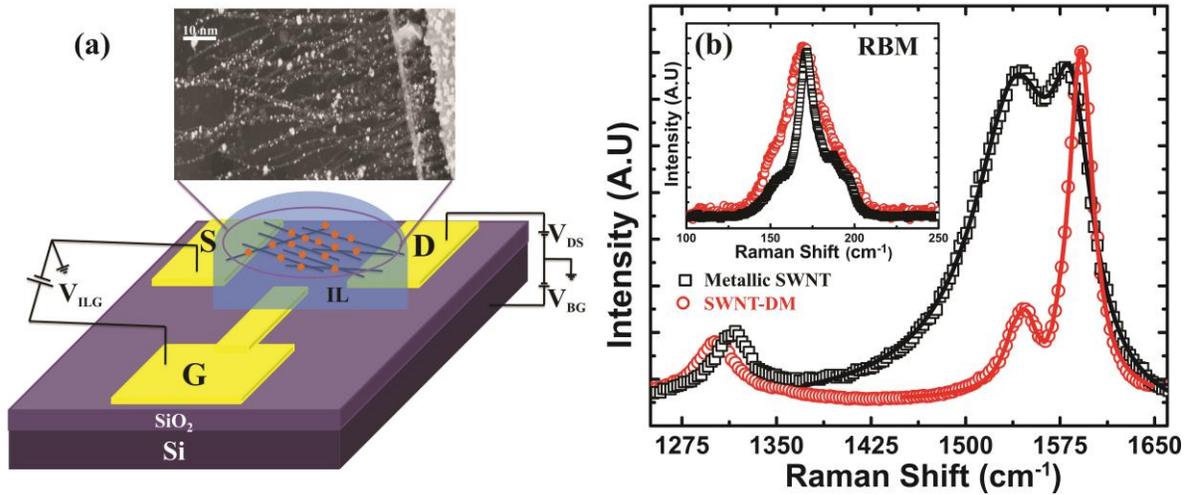

Figure 1: (a) Schematic of the FET device made of metallic SWNTs and inset shows the SEM image of DM molecules decorated metallic SWNTs. (b) Tangential mode of metallic SWNTs and SWNT-DM complex. Black colour open squares and solid line represent experimental data and BWF + Lorentzian fit for metallic SWNTs and red colour open circles and solid line represent experimental data and Lorentzian fit for SWNT-DM complex.

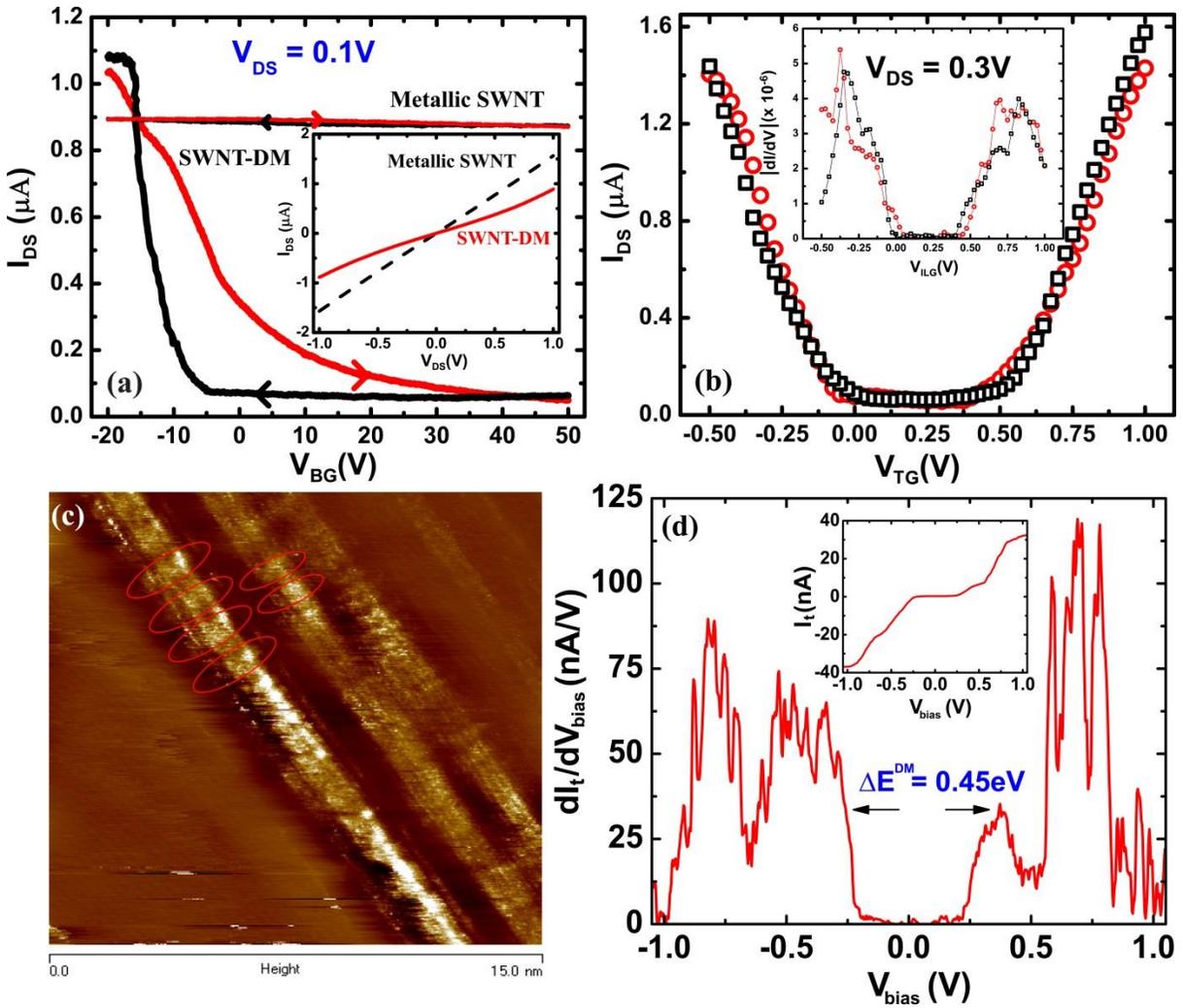

Figure 2: (a) Back gate transfer characteristics ($I_{DS}$ -$V_{BG}$) of the FET device made of metallic SWNTs before and after the complexation with DM molecules (Black colour and red colour solid lines shows forward and reverse runs). Inset shows $I_{DS}$ -$V_{DS}$ characteristics before (black colour dashed line) and after (red colour solid line) complexation. (b) Ionic liquid top gating transfer characteristics ($I_{DS}$ -$V_{TG}$) of the FET device made of SWNT-DM complex. Inset shows the derivative of $I_{DS}$ as a function of $V_{TG}$ (Red colour open circles are for the forward run and black colour open squares are for the reverse run). (c) Topographical image of the SWNT-DM complex. (d) Local dynamic tunnelling conductance as a function of bias voltage for the SWNT-DM complex (Inset: tunnelling current as a function of $V_{bias}$).

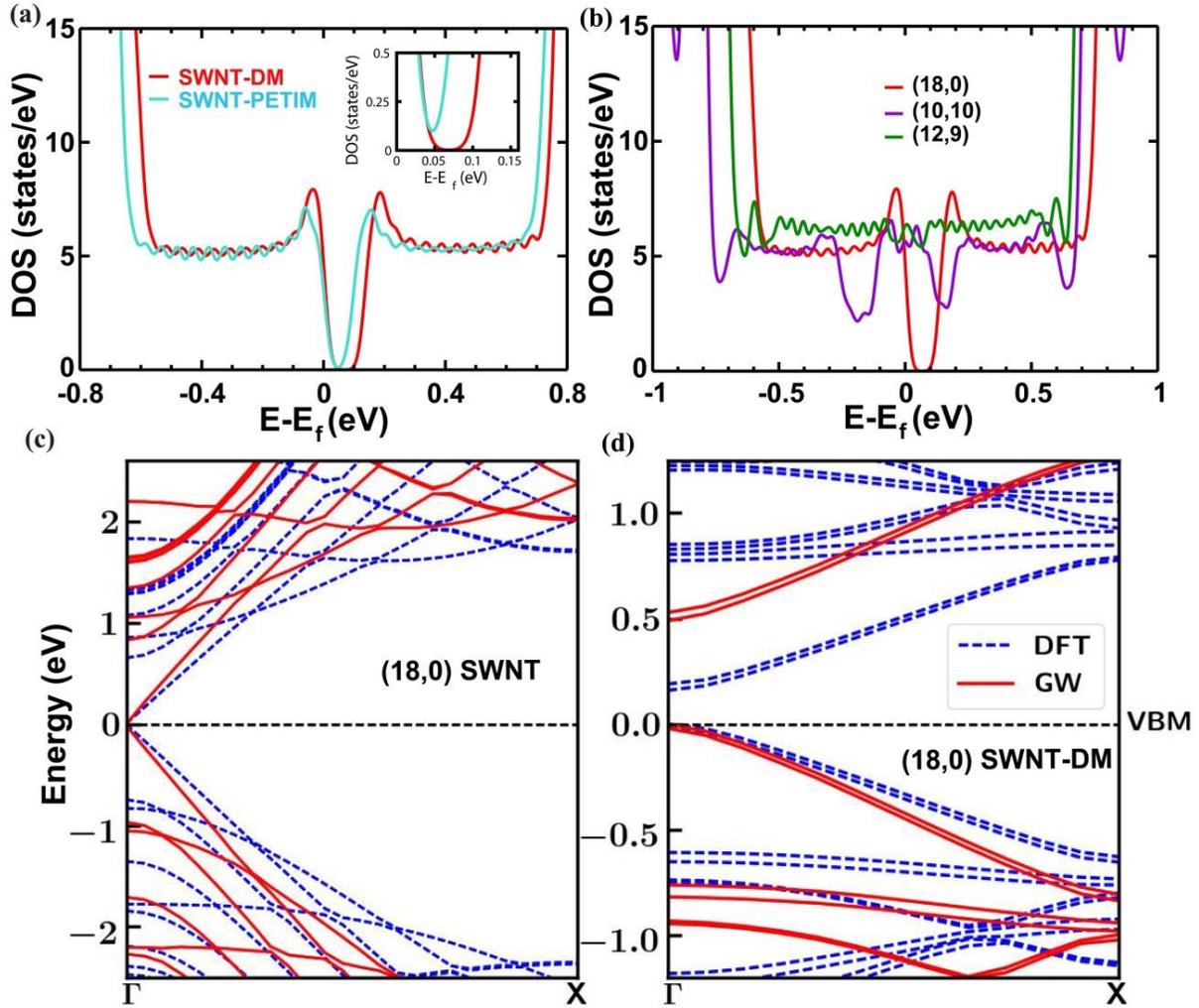

Figure 3: (a) Comparison of the DOS calculated for strained SWNT (18,0) wrapped by DM (red colour) and PETIM (cyan colour). DOS data has been averaged over 5 independent initial configurations for DM and 6 independent initial configurations for PETIM. Inset shows DOS near Fermi level. (b) Comparison in DOS calculated for strained SWNT wrapped by DM for three different chiralities, red (18,0), violet (10,10) and green (12,9) respectively. Statistical averages have been done over 5 different DOS calculations for (18,0) and over 10 different DOS calculations for (10,10), (12,9) respectively. Fermi level is shifted to zero in x-axis. 1 meV binwidth has been used with 25 meV Gaussian broadening. (c) Electronic band structure of pristine (18,0) SWNT and (d) SWNT-DM complex. Band structure of the (18,0) SWNT and DM

wrapped (18,0) SWNT is calculated within GW (red colour solid line) as well as DFT (blue colour dotted line).

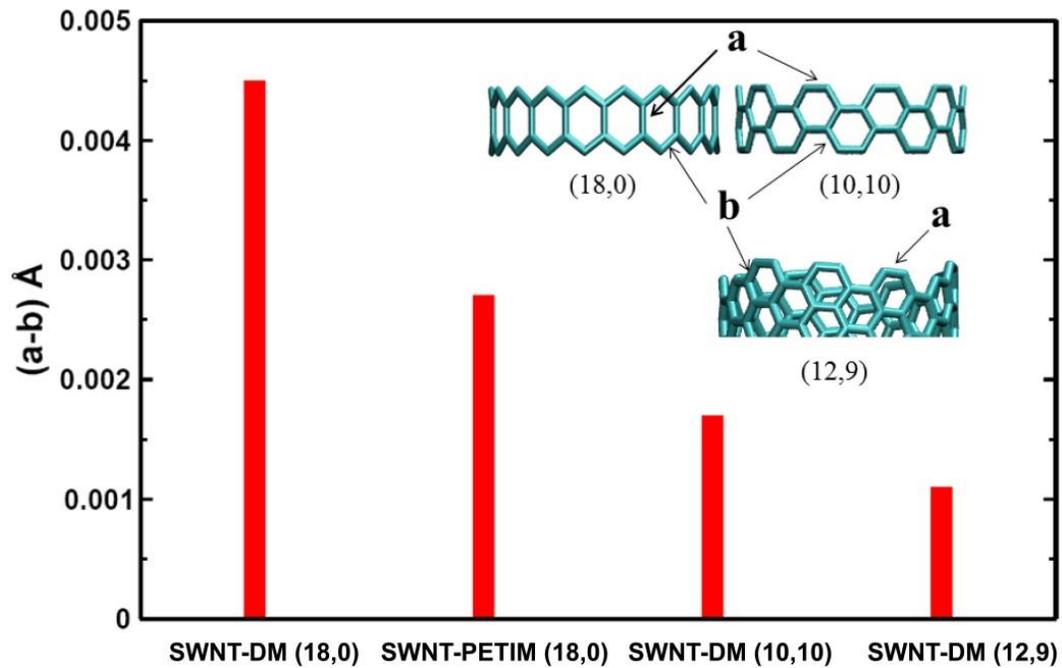

Figure 4: Plot showing a comparative measure of bond asymmetry [(a-b) in Å] between two types of bonds ['a' and 'b'] for SWNT-DM (18,0), SWNT-PETIM (18,0), SWNT-DM (10,10) and SWNT-DM (12,9), respectively.

# Opening of large band gap in metallic carbon nanotubes by mannose functionalized dendrimers: Experiments and theory

K. S. Vasu[a*], Debabrata Pramanik[a], Sudipta Kundu[a], Sridevi S[b], N. Jayaraman[c], Manish Jain[a], Prabal K. Maiti[a] and A. K. Sood[a]

[a]Department of Physics, Indian Institute of Science, Bangalore-560012, India.

[b]Department of Instrumentation and Applied Physics, Indian Institute of Science, Bangalore-560012, India.

[c]Department of Organic Chemistry, Indian Institute of Science, Bangalore-560012, India.

**Computational Details:**

We performed all atom MD simulations for SWNT and G4 DM along with water molecules at room temperature (T = 300 K) to get an equilibrium configuration of the composite structures. The AMBER14 software packages[1] were used for MD simulations with General Amber force field (GAFF)[2] to describe the intra and inter-molecular interactions. During 30 ns long MD simulations, DM molecule wraps around the SWNT surface to form composite structure, similar to SWNT-dendrimer structure for PETIM and PAMAM dendrimers[3,4]. This adsorption of dendrimer onto the SWNT surface produces a strain in the SWNT. Taking a portion of this strained SWNT (containing 5 unit cells along the tube axis), we calculate the electronic density of states (DOS) using density functional theory (DFT) as implemented in the Quantum Espresso software packages[5]. Periodic boundary conditions were used along nanotube axis to mimic infinite tube in DFT. Ultrasoft Pseudopotentials[6] were used to describe the electron-ion interactions. We used Perdew-Burke-Ernzerhof (PBE) approximation[7] to the exchange correlation functional. The wave functions (charge density) were expanded using plane waves[8] up to energy of 30 Ry (300 Ry). A 1x1x10 Monkhorst Pack k-point[9] sampling of the Brillouin zone with 0.005 Ry Fermi Dirac smearing, was used for self-consistent calculations. The k-grid was increased to $1\times1\times30$ for DOS calculation. To calculate the quasiparticle band gap in semiconducting CNTs, we performed GW calculations using Berkeley GW software package[10]. Norm-conserving pseudopotentials with PBE[7] exchange-correlation functional was used to calculate the mean-field wave-functions required for the GW calculations. The dielectric matrix was expanded in plane wave with energy up to 8 Ry, and was extended to finite frequency using generalized Plasmon pole (GPP) model[11]. The number of unoccupied bands used was 5 times occupied bands and $1\times1\times6$ k-grid was used for the GW calculations.

**RBM of metallic SWNTs with different laser excitations:**

The electronic transition energies ($E_{ii}$) are different for single walled carbon nanotubes (SWNTs) of different diameters. We have now recorded the radial breathing mode (RBM) of our metallic SWNTs samples with different laser excitations i.e. 532 nm, 633 nm and 660 nm. As shown in Fig. S2, RBM obtained with all three excitations has shown a high intensity band centered at 171 (±1) cm$^{-1}$, which corresponds to the diameter value of 1.45 (±0.01) nm. Besides this band, the RBM obtained with 660 nm excitation exhibited a moderately high intensity band centered at 161.1 cm$^{-1}$ (corresponds to a diameter of 1.55 nm) and the RBM from 633 nm excitation has shown other low intensity bands centered at 153.2, 187.4 and 196.7 cm$^{-1}$ (correspond to the diameter values of 1.63, 1.32 and 1.25 nm, respectively). Based on these observations, we confirm that our samples predominantly contain the SWNTs of diameter of ~ 1.45 nm with a partly contribution from the SWNTs of diameter of ~ 1.55 nm and a negligible contribution from the SWNTs of diameter of ~ 1.25 nm, 1.32 nm and 1.63 nm. Thus, one can envisage that the band gap opening discussed in this publication, will perhaps be feasible for the SWNTs of diameters between 1.4 to 1.6 nm.

**Band structure of metallic SWNTs and DM wrapped metallic SWNT:**

SWNT wrapped with DM molecule is taken to be periodic along the axial direction. As a result, the band structure is plotted only along the axial direction. In Fig. S7, we show the DFT band structure of pristine SWNT and DM wrapped SWNT (whereas Fig. 3c in the main manuscript shows band structure of pristine unit cell of SWNT). Although the bands look symmetric around the Fermi level, there is a significant rearrangement of the bands away from the Fermi level. Furthermore, several band degeneracies are lifted in the distorted nanotube.

**Bader charge analysis:**

In order to show that there is no charge transfer between the SWNT and DM, Bader charge analysis for the SWNT and DM composite has been performed (shown in Fig. S8). The quantitative value of the charge transferred between DM and SWNT is 0.0083 a.u. from the Bader charge analysis—indicating that there is no significant charge transfer.

**PDOS calculations and Fat band analysis:**

We have now performed PDOS calculations for the both pristine and distorted (18,0) SWNT. The contribution from the p-orbitals oriented along axial direction of the tube ($p_z$), radial direction of atoms ($p_r$) and circumferential direction of atoms ($p_\varphi$) to the DOS are calculated separately. Fig. S9(a) and (b) show PDOS plot for pristine and distorted tube respectively. We can see that only $p_r$ ($\pi$ orbital of the carbon atoms) contributes near the Fermi energy of the pristine and VBM of the distorted tube.

We have also performed the fat band analysis for the pristine and distorted tubes (Fig. S10a and S10b). The bands near the Fermi energy of the pristine tube or VBM of the distorted tube are of $p_r$ characteristic which is consistent with the PDOS calculation.

Thus from PDOS and the fat band analysis, it can be concluded that the band characteristic near the VBM of the distorted tube does not change when the tube is wrapped with DM.

**Back gate transfer characteristics:**

Fig. S11 illustrates the back gating characteristics of field effect transistor (FET) devices made of pristine metallic SWNTs (red color line) and metallic SWNTs functionalized with PETIM dendrimer molecules (black color line). It is clear that the behavior of gating characteristics of the FET device made of metallic SWNTs functionalized with PETIM dendrimer is similar to that of the FET device made of pristine metallic SWNTs. This confirms the absence of metal to semiconductor transition in metallic SWNTs functionalized with PETIM dendrimer molecules and hence PETIM dendrimer without mannose functionalization does not induce band gap opening in SWNTs. Thus, we are justified in confirming the observed band gap opening cannot be generalized for every dendrimer or DNA like supramolecules.

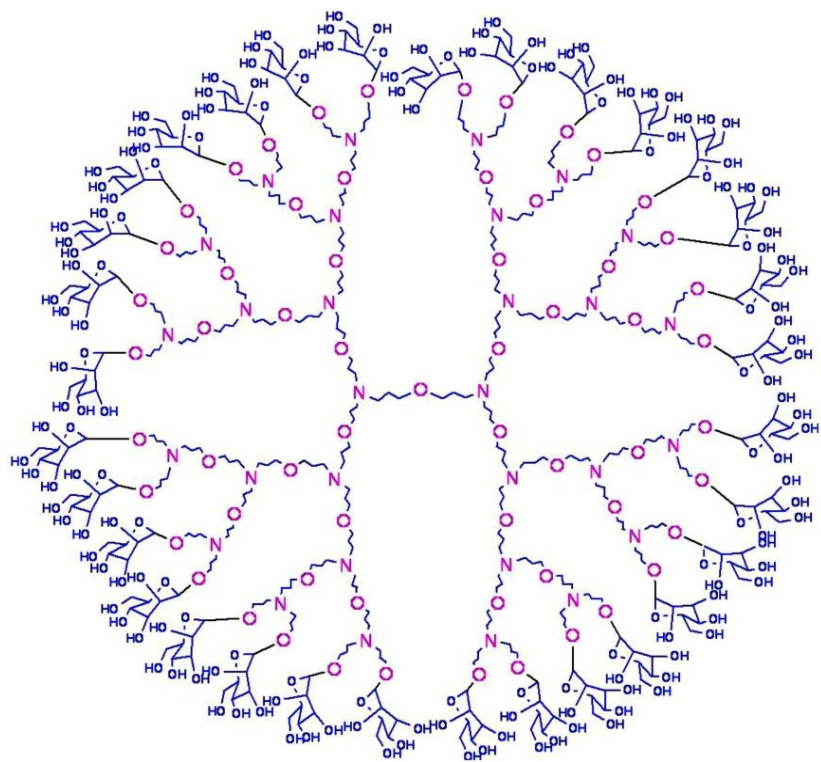

Figure S1: Chemical structure of mannose attached PETIM dendrimer.

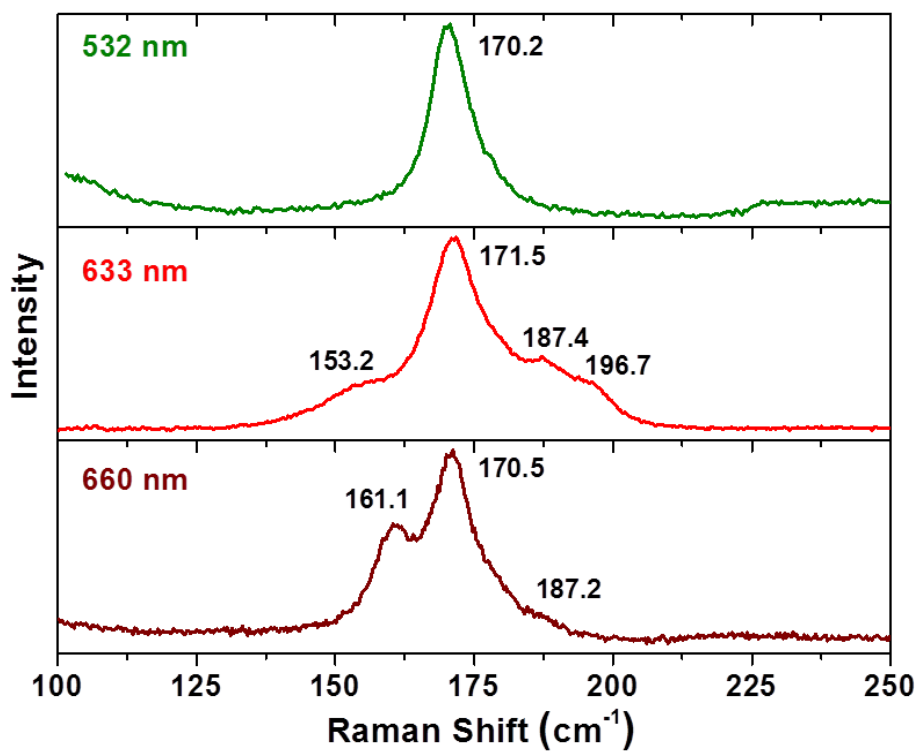

Figure S2: Radial breathing mode (RBM) of metallic SWNT samples with 532, 633 and 660 nm excitations.

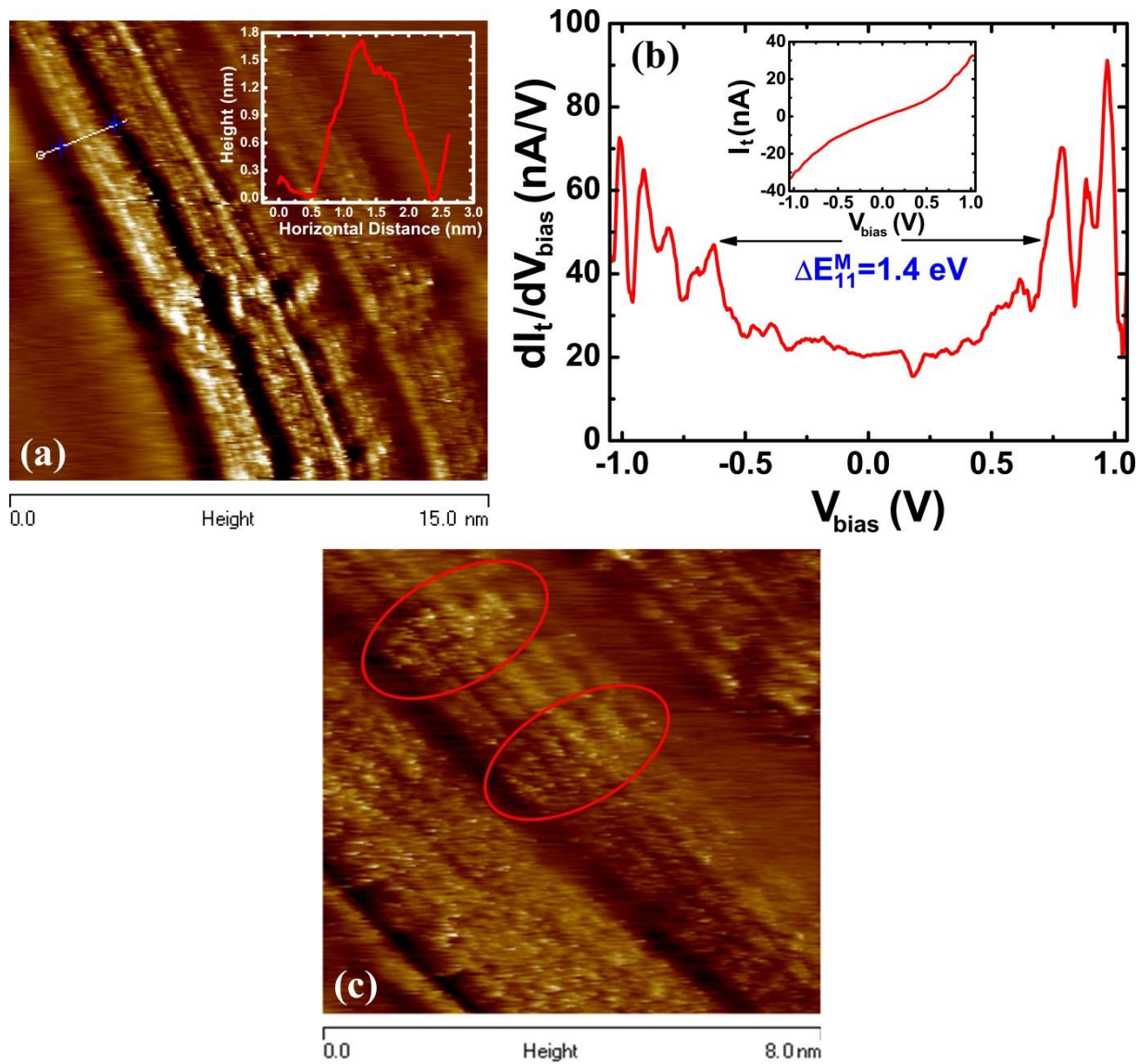

Figure S3: (a) Topographical image of metallic SWNTs (Inset: Height profile of metallic SWNTs). (b) Local dynamic tunnelling conductance as a function of bias voltage for metallic SWNT (Inset: tunnelling current as a function of $V_{bias}$). (c) High magnification topographical image of SWNT-DM complex on HOPG.

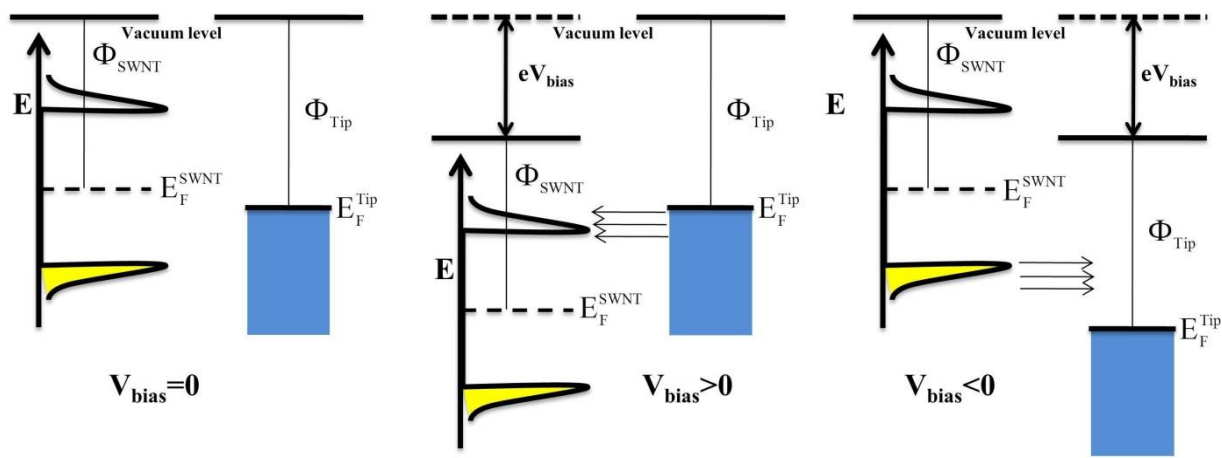

Figure S4: Energy level diagram explaining the electron transfer at $V_{bias} = 0$, positive and negative.

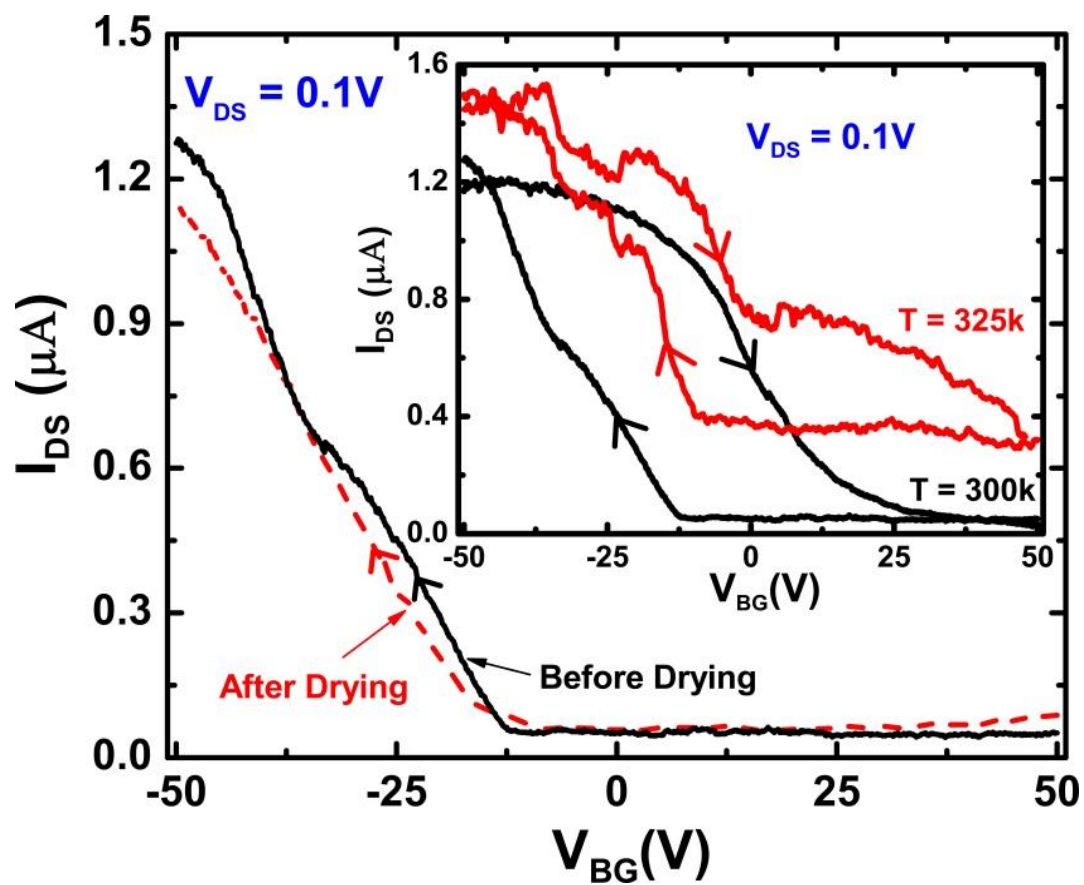

Figure S5: $I_{DS}$-$V_{BG}$ characteristics of another device made of SWNT-DM complex before (black colour solid line) and after (red colour dashed line) drying in vacuum and Inset shows $I_{DS}$-$V_{BG}$ characteristics of the same device at 300k and 325k.

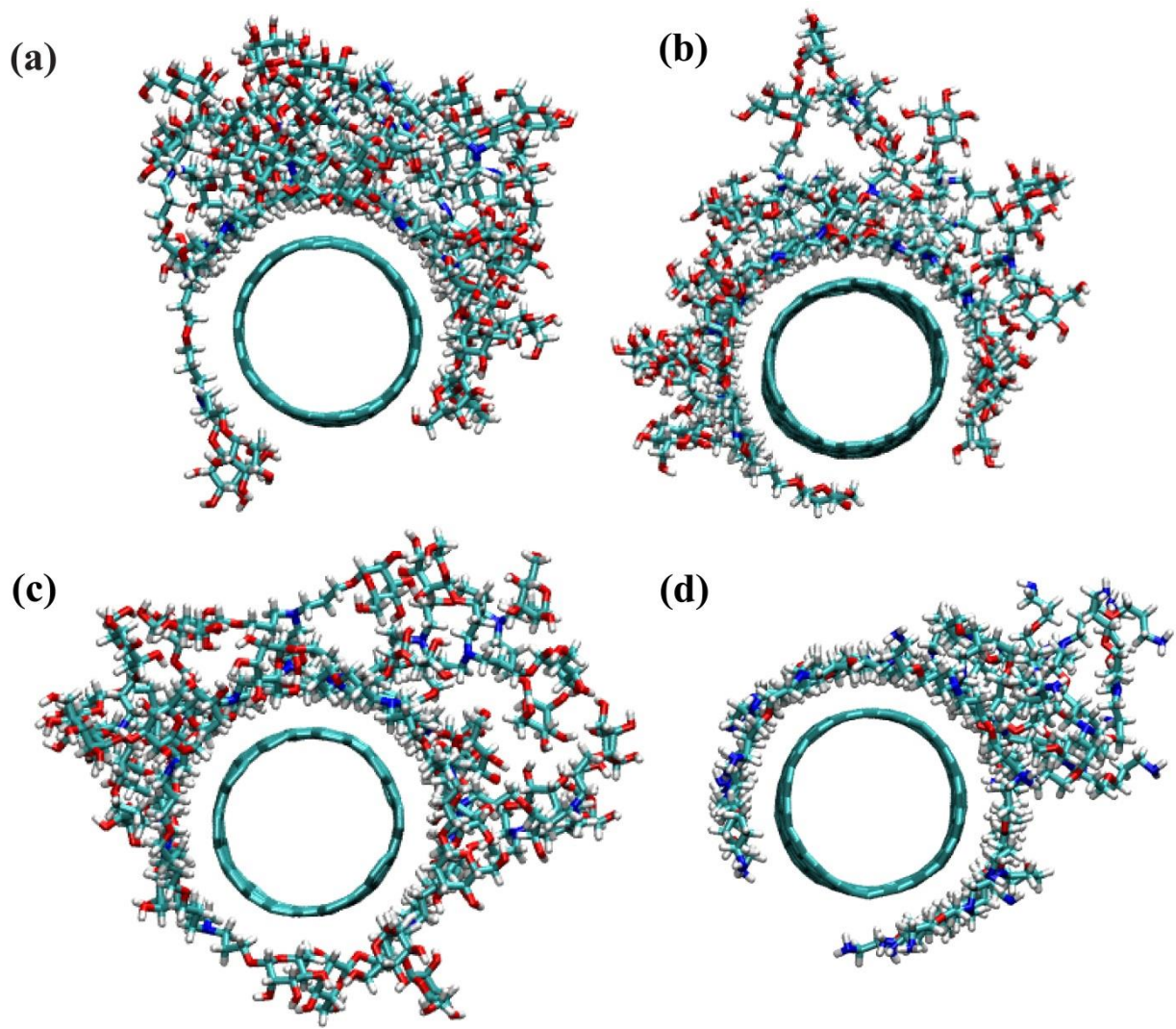

Figure S6: Instantaneous snapshots of equilibrium composite structures of dendrimer-NT after 30 ns of long MD simulations. Here the snapshots are of (a) (18,0) SWNT-DM (b) (12,9) SWNT-DM (C) (10,10) SWNT-DM and (d) (18,0) SWNT-PETIM. For clarity water molecules are not shown.

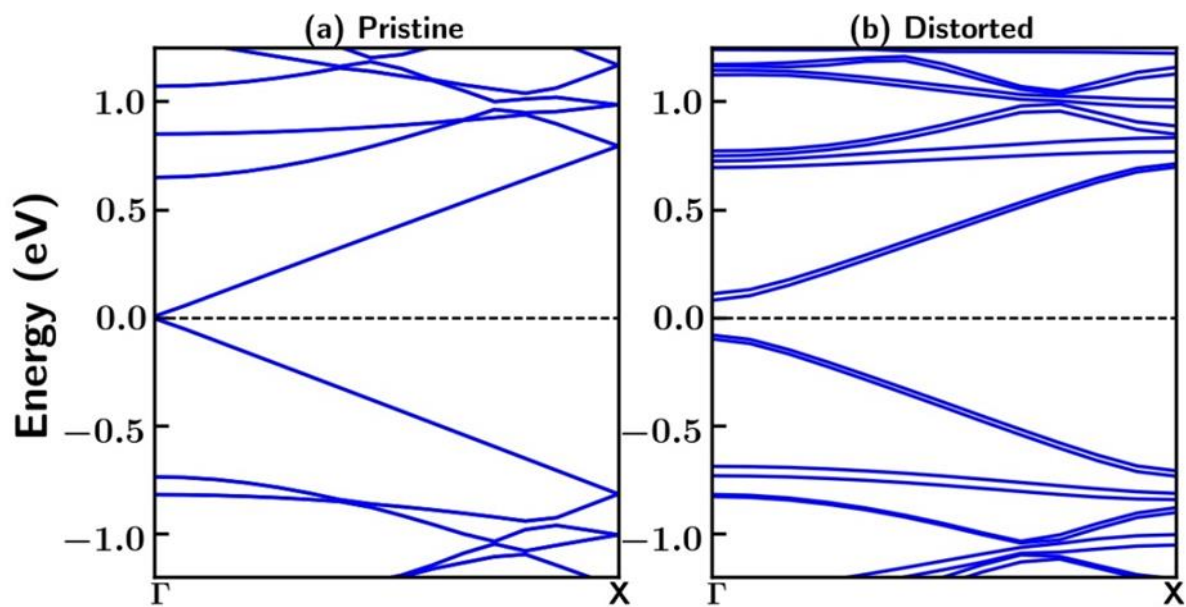

Figure S7: Band structure of the pristine (18,0) nanotube (5 unit cells along the nanotube axis) and corresponding DM wrapped nanotube. The bands are plotted along the nanotube axis.

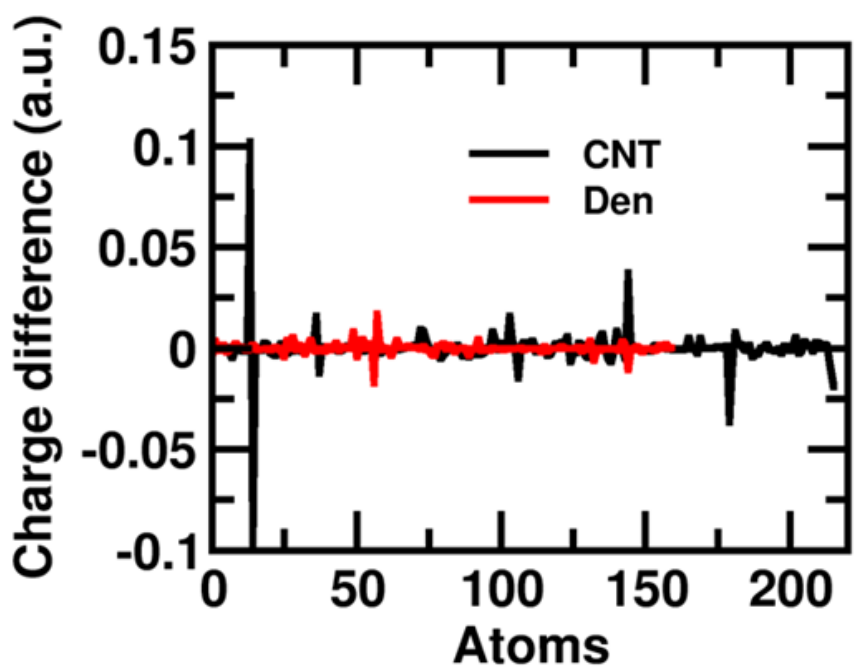

Figure S8: Bader charge analysis for SWNT and DM composite system. Charge difference between the isolated SWNT and SWNT of the composite is shown in black, whereas DM in red.

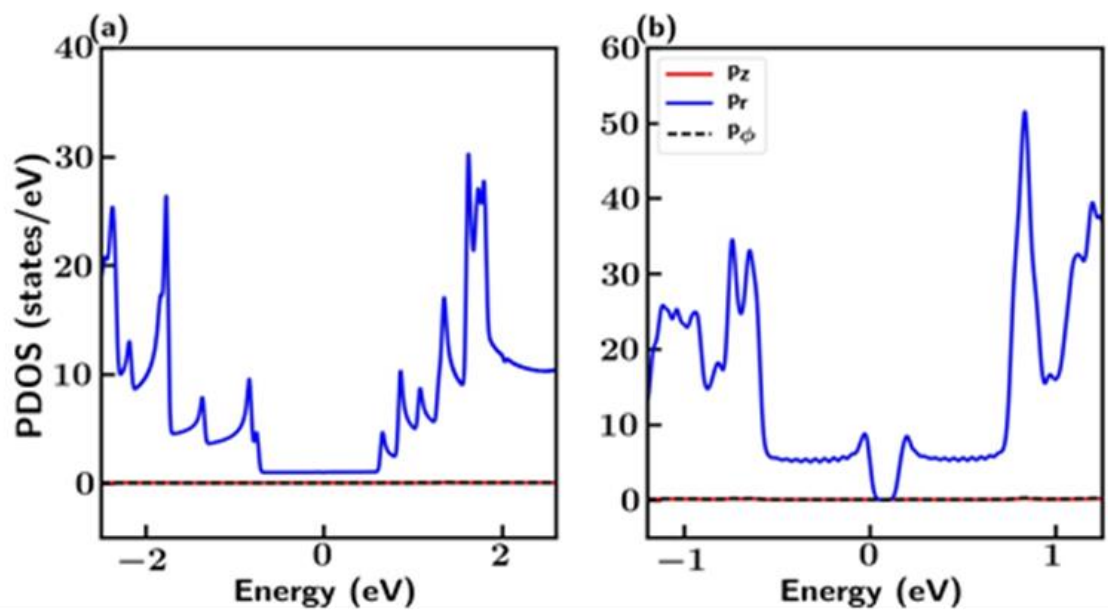

Figure S9: (a) and (b) shows PDOS for pristine and distorted tube respectively. The blue line shows the contribution from the $p_r$ orbitals and the red and black dotted line show the PDOS due to $p_z$ and $p_\varphi$. The zeros of x-axis are set to the Fermi energy for pristine and to VBM for distorted one.

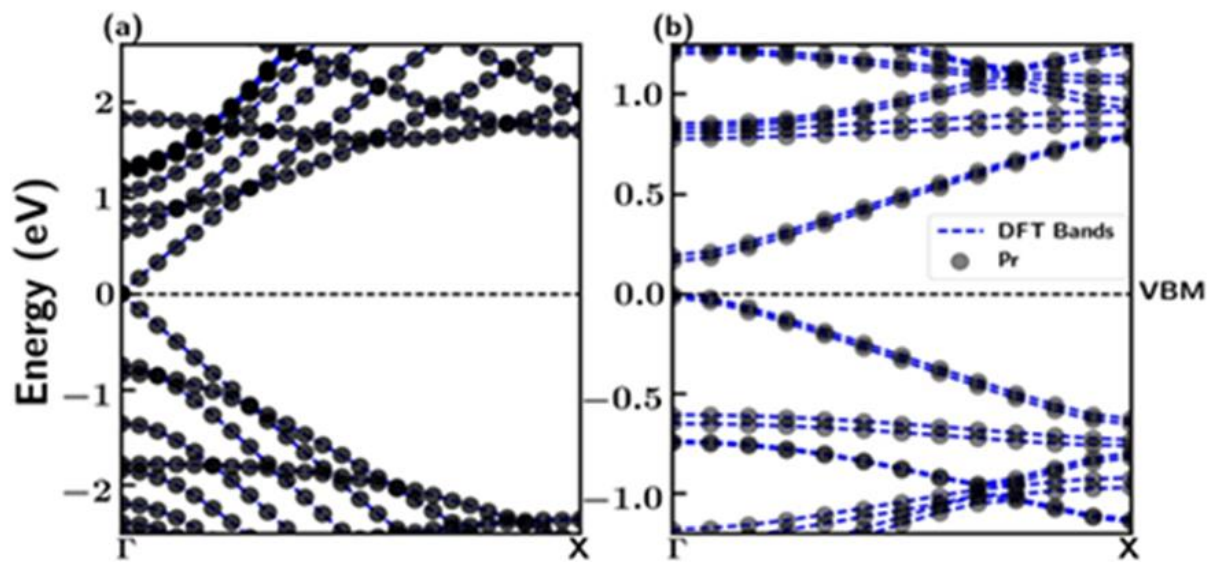

Figure S10: (a) and (b) show fat band for pristine and distorted tubes respectively. The blue dotted lines are bands obtained from DFT and the black dots give the contribution from $p_r$ orbital. The dots are of same size as all the bands in the energy window of the plots are totally composed of $p_r$ orbital.

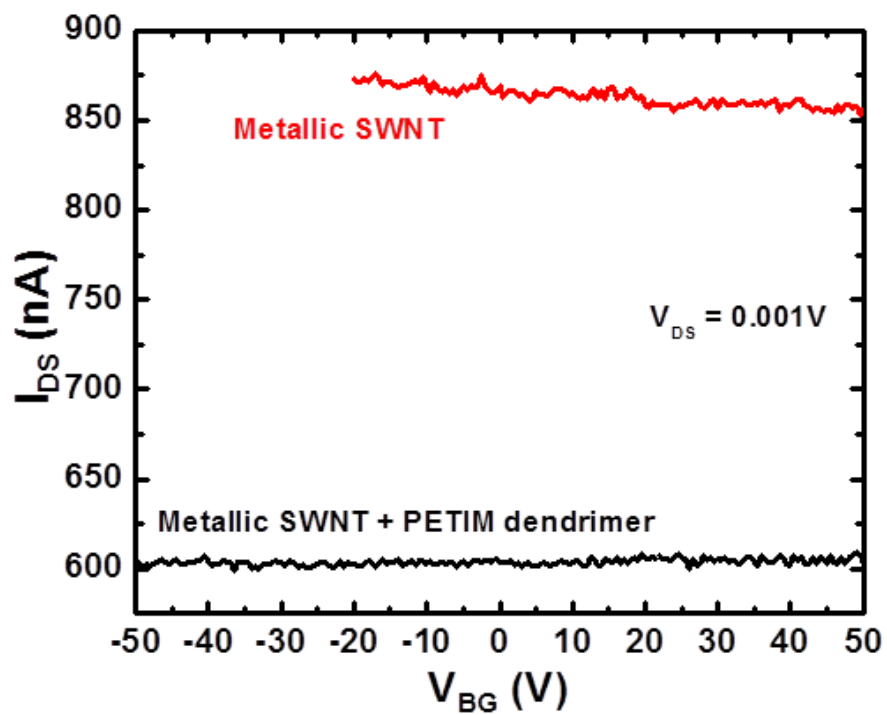

Figure S11: Back gate transfer characteristics ($I_{DS}$ -$V_{BG}$) of the FET device made of metallic SWNTs before (red colour solid line) and after the complexation with PETIM dendrimer molecules (Black colour solid line).